\documentstyle[12pt,epsfig,amsmath]{article}
\newcommand {\be}{\begin{equation}}
\newcommand {\ee}{\end{equation}}
\newcommand {\ba}{\begin{eqnarray}}
\newcommand {\ea}{\end{eqnarray}}
\begin{document}

\def \a'{\alpha'}
\baselineskip 0.65 cm
\begin{flushright}
%IPM/P-2007/000\\
\today
\end{flushright}
\begin{center}{\large
{\bf }}
 {\vskip 0.5 cm}
 {\bf Flavoring Monochromatic Neutrino Flux from Dark Matter
 Annihilation}

{\bf  Yasaman Farzan }{\vskip 0.5 cm }
School of physics, Institute for research in fundamental sciences (IPM)\\
P.O. Box 19395-5531, Tehran, Iran\\
\end{center}

\begin{abstract}
As is well-known, if the scattering cross section of the Dark
Matter (DM) particles off the nuclei inside the Sun is large
enough, the Sun can  trap DM particles. In principle, the
annihilation of DM pair inside the Sun can then lead to the
detection of  a relatively large flux of neutrinos in the neutrino
telescopes. If the annihilation directly produces a neutrino pair,
the flux of neutrinos on Earth will be monochromatic. In this
case, the oscillatory terms in the oscillation probability lead to
a novel seasonal variation of  detected events which is sensitive
to the initial flavor composition. In this paper, we propose two
models that predict such a detectable monochromatic neutrino flux
from the DM annihilation. Model I, which is based on augmenting
the type II seesaw  mechanism, predicts a  flavor composition for
the monochromatic flux determined by $(m_\nu)_{\alpha \beta}$. In
model II, the DM pair first annihilates to a pair of sterile
neutrinos which oscillate into active neutrinos with a flavor
composition determined by the flavor structure of the
active-sterile neutrino mixing.
\end{abstract}

\section{Introduction}
Although reasonably large amount of evidence has been accumulating
in favor of Dark Matter (DM) as the explanation of the missing
mass problem of the universe, the nature of DM particles is still
unknown. Under the assumption that the DM particles are produced
thermally in the early universe and there is only one kind of DM,
the measured DM abundance in the universe determines the
self-annihilation cross section to be 1~$p$b. However, the DM mass
or its annihilation products are still unknown.

Arguably one of the most plausible classes of DM is Weekly
Interacting Massive Particles (WIMP) with a mass, $m_{DM}$, in the
range $100~{\rm GeV}-{\rm few}$~TeV.   This range of DM mass is
interesting both for direct and indirect  techniques of DM
particle detection. Direct detection techniques are based on
measuring the recoil energy in the scattering of DM particles off
nuclei in a target. Indirect detection techniques are based on
detecting the particles produced by DM annihilation in regions
such as the  Sun  or the galaxy center where the DM concentration
is relatively high.

Within certain popular models such as MSSM, the DM pairs
 first annihilate to pairs such as $b \bar{b}$,
$\tau \bar{\tau}$, $Z {Z}$ and $W^+W^-$. Neutrinos are then
produced as secondaries through their decays.  However, as
discussed in \cite{Manfred}, there are various ways to build a
model in which DM particles dominantly annihilate to neutrinos.

 In this paper, we are  interested in detecting the
neutrino flux from the annihilation of DM particles that have been
accumulated inside the Sun. These DM particles are
non-relativistic. As a result, if neutrinos are directly produced
by the DM pair annihilation, the spectrum will be monochromatic
with $E_\nu=m_{DM}$. However, if neutrinos are secondary products
of the DM annihilation, their spectrum will be continuous with
$E_\nu<m_{DM}$. In either case, if the DM mass, $m_{DM}$, is
larger than the detection threshold of the neutrino telescopes, a
high energy neutrino flux pointing towards the Sun is expected at
ICECUBE. Remember that the energy of ordinary neutrinos produced
by proton fusion in the Sun center is too small to be detectable
by neutrino telescopes with detection energy threshold of
$E_{th}\gg$a few 10 MeV.

 The purpose
of the present paper is to  build  models within which the method
proposed in \cite{Esmaili1,Esmaili2} is effective. Such a model
should have
 the following properties: (1) The Dark matter
mass is larger than the detection energy threshold of  the
neutrino telescope: O(50~GeV)-O(100~GeV). (2) The DM pair
dominantly annihilates to a neutrino pair with a non-trivial
flavor composition and a total annihilation cross section of
1~$p$b. (3) As it will be discussed in the next section, to obtain
enough statistics, the scattering cross section of DM particles
off nucleons should be greater than $10^{-9}~p$b. On the other
hand, it should be smaller than $10^{-8}~p$b to evade the bounds
from direct detection \cite{direct}. Other examples of such models
can be found in Ref. \cite{example}.

In sect. \ref{review}, we briefly review the method proposed in
\cite{Esmaili1,Esmaili2} and formulate the conditions a model has
to satisfy to predict a sizable seasonal variation in the number
of events from the DM annihilation in the Sun center. In sect.
\ref{ModelI}, we propose Model I  which embeds type II seesaw
mechanism. Within this model, the flavor structure of $\sigma({\rm
DM}+{\rm DM} \to \nu_\alpha \nu_\beta)$ is determined by
$(m_\nu)_{\alpha \beta}$. In sect. \ref{modelII}, we propose Model
II  within which the  DM pair first annihilates to a pair of
sterile neutrinos and then the sterile neutrinos oscillate into
active neutrinos on the way to Earth. In sect. \ref{con}, we
review our conclusions.

\section{A novel method to extract information about dark matter
particles \label{review}}

The DM particles propagating in the solar system have velocities
about a few hundred km/sec. When these particles enter the Sun,
they can lose their kinetic energy by scattering off the nuclei.
They will then fall in the gravitational well of the Sun. As a
result, during the Sun lifetime, the density of DM in the Sun has
increased. The number density increases with the scattering cross
section of these particles off nuclei. The trapped DM particles
virilize and  come to thermal equilibrium with the nuclei in the
Sun center. Equating $|E_{kinetic}|=3 k_B T_\odot/2$ with
$|V_{gravity}|=4\pi G_N\rho m_{DM}r_{DM}^2/3$, we find that the
virilized DM particles are centered in the Sun within a volume of
radius $r_{DM}\sim (9 k_BT_\odot / 8\pi G_N \rho_\odot
m_{DM})^{1/2}$. The DM particles inside the Sun are
non-relativistic so the annihilation of a DM pair into a pair of
on-shell particles will result in a  monochromatic spectrum. In
case that a neutrino pair is directly produced by the annihilation
of the DM pair, the energy of each neutrino will be equal to the
DM mass, $m_{DM}$. However, if the neutrinos are the decay
products of  unstable particles  produced by the DM pair
annihilation, their spectrum will be continuous. In the latter
case because of the very large distance between the Sun and Earth,
$L_{Sun-Earth}$, the oscillatory terms in neutrino oscillation
probability given by $\sin( \Delta m^2 L_{Sun-Earth}/2E_\nu)$ will
average to zero. However, as discussed in \cite{Esmaili1}, in the
former case where the spectrum is monochromatic, the oscillatory
effects in the oscillation probability are not averaged out and
therefore lead to a seasonal variation in the number of events at
ICECUBE as the Sun-Earth distance varies during a year because of
the eccentricity of the Earth orbit. In \cite{Esmaili2}, it was
shown that studying this seasonal variation provides information
on the flavor structure of ${\rm DM}+{\rm DM}\to
\nu_\alpha+\stackrel{(-)}\nu_\beta$ as well as on the value of
$m_{DM}$ (through the combination $\Delta m_{21}^2/m_{DM}$). In
particular, observing oscillatory behavior on top of the trivial
variation of inverse of the square of  the Earth Sun distance
({\it i.e.,} $1/L_{Earth-Sun}^2$) means the initial flux is
monochromatic with a non-trivial non-democratic flavor composition
({\it i.e.,} $F_{\nu_e}:F_{\nu_\mu}:F_{\nu_\tau}\ne 1:1:1$).

 In  \cite{Esmaili2}, the details of this novel method to extract
 information on the properties of DM annihilation modes have been
 discussed. % As emphasized before measuring the effects of the
 %oscillatory terms requires two conditions: (1) direct
 %annihilation of the DM pair to neutrinos and therefore a
 %monochromatic spectrum; (2) non-trivial flavor composition at
 %production.
   Of course, to
study the time variation of the neutrino flux, sufficient
statistics is also required. The statistics is determined by the
rate at which the Sun traps DM particles and this rate in turn
depends on the scattering cross section of the DM particles off
the nuclei in the Sun. Scattering of DM particles off nucleons
inside a nucleus is coherent so in the case that scattering is
spin independent, the cross section of DM particles off a nucleus
of $Z$ protons and $A-Z$ neutrons is proportional to $|Z
{\mathcal{M}}_p+(A-Z) {\mathcal{M}}_n|^2$ where ${\mathcal{M}}_p$
and $ {\mathcal{M}}_n$ are respectively the scattering amplitude
of  the DM particles off a single proton and neutron. (Notice that
if the scattering was not coherent, the cross section would be
given by $Z |{\mathcal{M}}_p|^2+(A-Z)|{\mathcal{M}}_n|^2$.)
Because of the non-linear dependence on $Z$ and $A-Z$, heavier
nuclei can trap the DM more effectively than the same number of
separate nucleons. Thus, although the majority of the mass of the
Sun is composed of protons, in evaluating the capture rate via
spin-independent scattering, the presence of the heavier nuclei
inside the Sun has to be taken into account.  It can be shown that
for $m_{DM}>100$ GeV, if the spin-independent cross section is
larger than $10^{-9}$~pb and each DM pair produces a neutrino
pair, a few hundred events can be registered each year in a
detector such as ICECUBE \cite{Esmaili2,jungman}. In this mass
range, the direct bounds on the cross-section of the
spin-independent scattering of DM particles off nucleons is $\sim
10^{-8}$ pb.

In \cite{Esmaili1}, various possible sources of widening of the
monochromatic spectrum from ${\rm DM}+{\rm DM}\to
\nu_\alpha+\nu_\beta$ have been studied. The main source of
widening turns out to be the thermal distribution of the
velocities of the initial DM particles. This thermal distribution
widens the monochromatic line to a narrow Gaussian with a width
$\Delta E/E\sim 10^{-4}(100~{\rm GeV}/m_{DM})^{1/2}$
\cite{Esmaili1}. At these energies a fraction of active neutrinos
undergo interaction with the nuclei inside the Sun while they
cross the Sun. The charged current  interactions of
$\stackrel{(-)}{\nu}_e$ and $\stackrel{(-)}{\nu}_\mu$ produce
charged leptons which are absorbed.
 The
charged current  interaction of $\stackrel{(-)}{\nu}_\tau$
produces a charged tau which in turn decays  producing
$\stackrel{(-)}{\nu}_\tau$ with a lower energy. The neutral
current interaction of the neutrinos produces a neutrino of a
lower energy. The outcome is that the scattering will reduce the
height of the monochromatic line but will add a tail to the
spectrum. Since the cross section of the neutral current
interaction is finite in the forward scattering, the sharp line
will remain sharp.  At the sun surface the spectrum will be
composed of a sharp line superimposed on the end of a continuous
spectrum consisting of the scattered and regenerated neutrinos.
For $m_{DM}<500$ GeV, the energy and therefore cross section of
the produced neutrinos are small which means a  good fraction of
the neutrino will remain unscattered resulting in a significant
seasonal variation due to the oscillatory terms in the oscillation
probability. However, for larger $m_{DM}$, the majority of the
neutrinos are supposed to be scattered before leaving the Sun.
This means in general for $m_{DM}>500$ GeV, the method proposed in
\cite{Esmaili2} is ineffective. In building Model I, we take this
consideration into account. However, in the case of Model II, DM
pair first annihilates into sterile neutrinos that leave the Sun
unscattered. These neutrinos then oscillate to active neutrinos.
This means in the case of Model II, even if $m_{DM}>500$ GeV, the
method might be  effective.

On the other hand, if $m_{DM}<{\rm a ~few}\times 10$~GeV, the
energy of neutrinos will be below the detection threshold of
ICECUBE. Moreover, for lower values of the DM mass, the
correlation between the direction of incoming neutrinos and the
produced charged lepton will be lost. This in turn means using the
directionality to reduce the background from the atmospheric
neutrinos will become less efficient. Considering these facts, in
model building we set $m_{DM}>100$~GeV.

%%%%%%%%%%%%%%%%%%%%%%%%%%%
%%%%%%%%%%%%%%%%%%%%%%%%%%%%
%%%%%%%%%%%%%%%%%%%%%%%%%
\section{MODEL I \label{ModelI}}
%%%%%%%%%%%%%%%%%%%%%%%%
%%%%%%%%%%%%%%%%%%%%%%%%5
%%%%%%%%%%%%%%%%%%%%%
In this section, we build a  model for neutrino mass and dark
matter by slightly augmenting  the type II seesaw mechanism.
Through the standard type II seesaw mechanism, this model will
lead to a Majorana mass for neutrinos. The DM candidate in this
model can be either a complex scalar or a Dirac fermion that we
add to the model. A $Z_2$ symmetry guarantees its stability.
% As we
%shall see, the DM pair dominantly annihilates to a neutrino pair
%with a non-democratic flavor composition given by neutrino mass
%matrix elements. As we shall see, the DM nucleon scattering cross
%section can be large enough to lead to sufficiently large DM
%capture rate inside the Sun. In other words, the scattering cross
%section of the DM particles off nuclei can be larger than
%$10^{-9}$ pb.

As is well-known, within the type II seesaw mechanism,  a $SU(2)$
triplet scalar exists with a nonzero hypercharge as follows
\ba\Delta=\left[
\begin{matrix} {\Delta^+\over\sqrt{2}} & \Delta^{++} \cr \Delta^0
& - \frac{\Delta^+}{\sqrt{2}}
\end{matrix}\right]\ . \ea
The left-handed leptons couple to this triplet as follows \be
\label{fAB} {\mathcal{L}}= \frac{f_{\alpha \beta}}{2}\epsilon_{ik}
\bar{L}^c_{\alpha i } \Delta_{kj} L_{\beta j}=\frac{f_{\alpha
\beta}}{2}\left(\Delta^0
\nu_\alpha^T{\mathcal{C}}\nu_\beta+\frac{\Delta^+}{\sqrt{2}}(\nu_\alpha^T
{\mathcal{C}} l_\beta +l_\alpha^T {\mathcal{C}}
\nu_\beta)+\Delta^{++} l_\alpha^T{\mathcal{C}}l_\beta \right)\ee
in which $\alpha$ and $\beta$ are flavor indices, $i$, $j$ and $k$
are $SU(2)$ indices and ${\mathcal{C}}$ is the $2\times 2$ charge
conjugation matrix: ${\mathcal{C}}_{11}={\mathcal{C}}_{22}=0$ and
${\mathcal{C}}_{12}=-{\mathcal{C}}_{21}=1$. By assigning lepton
number equal to $-2$, the lepton number will be preserved by the
coupling in Eq.~(\ref{fAB}). The pure scalar part of the potential
involving only $H$ and $\Delta$ is given by \be
\label{potentialII} V=m^2 H^\dagger \cdot H+M_\Delta^2
Tr[\Delta^\dagger \Delta] +\frac{\lambda_1}{4} (H^\dagger \cdot
H)^2+\lambda_2H^\dagger \Delta^\dagger \Delta
H+\lambda_3Tr[\Delta^\dagger \Delta]H^\dagger \cdot H\ee
$$+\frac{\lambda_{\Delta 1}}{4} \left( Tr[\Delta^\dagger
\Delta]\right)^2+\frac{\lambda_{\Delta 2}}{4} \left| Tr[\Delta
\Delta]\right|^2  $$
$$+ \mu\left((\Delta^0)^\dagger
(H^0)^2+{\sqrt{2}}{\Delta^-}H^+ H^0+\Delta^{--} H^+H^+\right)\ .$$
Notice that other forms of quartic coupling of $H$ and $\Delta$
which preserves lepton number and electroweak symmetry can be
rewritten as the combinations of the above terms. For example,
${\rm Tr}(\Delta \Delta^\dagger [\Delta^\dagger ,
\Delta])=|Tr(\Delta \Delta)|^2-[Tr(\Delta^\dagger \Delta)]^2$,
$Tr(\Delta^\dagger \Delta \Delta^\dagger
\Delta)=(Tr(\Delta^\dagger\Delta))^2-|Tr(\Delta\Delta)|^2/2$ or
$H^\dagger \Delta\Delta^\dagger H=Tr(H^\dagger H)Tr(\Delta^\dagger
\Delta)-H^\dagger \Delta^\dagger \Delta H$. If
 the quartic couplings are all positive the potential will be stable; {\it i.e.,}
as $\Delta$ and/or $H \to \infty$, $V$ remains positive. In fact,
a weaker condition guarantees stability: For example, as long as
$\lambda_{\Delta 1}>0$, the  condition $\lambda_{\Delta
1}+\lambda_{\Delta 2}>0$ guarantees stability even if
$\lambda_{\Delta 2}$ is negative.

  The terms in the
third line of Eq.~(\ref{potentialII}) can be  rewritten as
$\epsilon_{ik} H_{ i } \Delta_{kj}^\dagger H_{ j}$. Since we have
assigned lepton number of $-2$ to $\Delta$, lepton number will be
broken only softly by $\mu$. Along with $H$, $\Delta^0$ also
receives a tiny VEV proportional to the lepton number violating
$\mu$ parameter \be \langle H\rangle =\frac{v}{\sqrt{2}} \ \ \ \
\langle\Delta^0 \rangle =\frac{-\mu v^2}{2M_\Delta^2+\lambda_3
v^2}\ .\label{VEVs}\ee Since $\mu v/m_\Delta^2 \ll 1$, we expect
$\langle\Delta^0 \rangle \ll \langle H\rangle$.  The first term in
Eq.~(\ref{fAB}) then gives a Majorana mass to neutrinos
$$(m_\nu)_{\alpha \beta}= -f_{\alpha \beta}\langle\Delta^0
\rangle=f_{\alpha \beta}\frac{\mu v^2}{2M_\Delta^2+\lambda_3 v^2}\
.$$

%By assigning lepton number of -2 to $\Delta$,
Although the coupling in Eq.~(\ref{fAB}) preserves the total
lepton number, it violates
 lepton flavor and can therefore give a significant contribution
 to Lepton Flavor Violating (LFV) processes. For a comprehensive
 review see \cite{Abada}. In particular, the bounds on $\mu \to
 eee$ sets the bound $f_{ee}f_{\mu e}/m_\Delta^2<1.2\times 10^{-5}~{\rm TeV}^{-2}$.
 %Let us
 %suppose $f_{ee}$ is much smaller than the rest of $f_{\alpha
 %\beta}$. From the bounds on LFV rare decay, we expect
 %$f_{\alpha\beta}\stackrel{<}{\sim} 0.1 (m_\Delta/1~{\rm TeV})^2$.
Taking $m_\Delta\sim 1~$TeV and $f_{\alpha \beta}\sim 0.01-0.001$,
all the bounds will be satisfied.

 For
$m_\Delta \sim $ TeV and $f_{\alpha \beta}\sim {\rm few}\times
10^{-3}$, the values of $\mu$ smaller than 10~keV result in small
enough neutrino mass.
 The smallness of the $\mu$ parameter can be justified by 'tHooft
 criterion: In the limit that $\mu$ vanishes, the lepton number is
 preserved.

Let us now discuss the DM sector. In the following, we discuss two
kinds of candidates: (1) complex scalar; (2) Dirac Fermions. In
both cases, we assign a lepton number equal to 1 to the DM
candidate. We introduce a $Z_2$ symmetry under which only the DM
candidate is odd. The $Z_2$ symmetry stabilize the DM candidate.
In order for the DM pair to annihilate to neutrinos, we need to
introduce another scalar, $\eta$, which mixes with $\Delta^0$. A
similar model in the context of extra dimensions has been studied
in \cite{Ma}. Before discussing the DM particles, let us focus on
$\eta$.
 We take $\eta$ to be  a singlet complex scalar with lepton number
 opposite to that of
 $\Delta$ with the following interaction term
 $$V=\lambda_4 \eta \epsilon_{jk}H^*_i \Delta_{ik} H^*_j$$
 which after electroweak symmetry breaking leads to a mixing term:
 $$ \lambda_4 \frac{v^2}{2} \eta \Delta^0 \ .$$
 Taking $m_\Delta^2 \gg m_\eta^2\sim \lambda_4 v^2/2$, we shall
 have two mass eigenstates with masses approximately equal to
 $m_\eta^2$ and $m_\Delta^2$ and mixing of $\lambda_4
 v^2/(2m_\Delta^2)$.

 A term such as $\eta^2 H^\dagger\cdot H$ is a  lepton number
 breaking term with a dimensionless coupling
so we do not include it.  We should protect $\eta$ from getting a
large VEV; otherwise it will lead
 to a large $\langle\Delta^0 \rangle$ and therefore large neutrino
 mass.  A term of  form \be \label{muPrime}\mu^\prime \eta
H^\dagger\cdot H\ee   leads to a mixing between $\eta$ and $h$
given by $\mu^\prime v/(m_h^2-m_\eta^2)$ and induces $\langle
\eta\rangle =-\mu^\prime v^2/m_\eta^2$. Notice that $\mu^\prime$
breaks lepton number softly, so it should be also suppressed:
$\mu^\prime \ll m_\eta$. The subsequent shift of $\Delta^0$ will
be given by $\lambda_4 (\mu^\prime)^2 v^2/m_\eta^2$. This small
shift in $\langle \Delta^0\rangle$ does not change the situation.
A small lepton number violating mass term of form $\eta^2$ can be
also added to the Lagrangian but it has no serious impact on the
discussion.

 As mentioned above, the DM can be either a complex scalar or a
 Dirac fermion.
 In both cases, the DM pair annihilates to a neutrino
pair via a $s$-channel exchange of a scalar mass eigenstates that
are mixtures of $\eta$ and $\Delta^0$. Let us now discuss both
possibilities one by one.

\begin{itemize}
\item\textbf{Complex scalar, $\Phi$, as DM}

The general $Z_2$ invariant and lepton number conserving
Lagrangian involving $\Phi$ can be written as \be {m_{\Phi}^2}
\Phi^\dagger \cdot\Phi+(\frac{m_{\eta \Phi \Phi}}{2}\eta \Phi
\Phi+ {\rm H.c.})+\ee
$$\frac{\lambda_\Phi}{4}(\Phi^\dagger \cdot\Phi)^2+{\lambda_{H\Phi}} H^\dagger\cdot H \Phi^\dagger
\Phi+ \lambda_{\eta \Phi} \eta^\dagger \eta\Phi^\dagger
\cdot\Phi+\lambda_{\Delta \Phi} {\rm Tr}(\Delta^\dagger
\Delta)\Phi^\dagger \cdot\Phi\ .$$ The coupling in the last line
mixes  $\eta$ and $\Delta$ so leads to
$$\sigma({\rm DM}+{\rm DM}  \to  \nu_\alpha \nu_\beta) =
\frac{m_{\eta \Phi \Phi}^2}{32 \pi }\left(\frac{\lambda_4 v^2
f_{\alpha \beta}}{[m_\eta^2-(2 m_{DM})^2][m_\Delta^2-(2
m_{DM})^2]}\right)^2\ ,$$ where $m_{DM}^2=m_\Phi^2+\lambda_{H\Phi}
v^2/2$. To obtain annihilation rate  indicated by DM abundance in
the standard thermal DM scenario, $\sigma_{tot} \sim 10^{-36}~{\rm
cm}^2$, the following relation should hold \be \left|
m_\eta^2-(2m_{DM})^2\right|\sim 300~{\rm
GeV}^2~\frac{\lambda_4m_{\eta \Phi \Phi}}{500~{\rm GeV}}
\frac{f_{\alpha \beta}}{0.01} \left(\frac{1~{\rm
TeV}}{m_\Delta}\right)^2\ . \label{split}\ee As mentioned earlier,
a $Z_2$ symmetry stabilizes $\Phi$ against decay. If $\langle
\Phi\rangle$ is nonzero, the $Z_2$ symmetry will be broken and
various decay modes will become open for $\Phi$. Vanishing
$\langle \Phi\rangle$ sets bounds on the parameters of the model.
A necessary condition for vanishing $\langle \Phi\rangle$ is \be
m_{\eta \Phi \Phi}^2<4(4\lambda_{\eta
\Phi}+\lambda_{\eta}+\lambda_{ \Phi})(m_{DM}^2+m_\eta^2)\
,\label{bbb}\ee where $\lambda_\eta$ is the quartic coupling of
$\eta$. If this bound is not satisfied, the minimum of the
potential will lie at
$$\Phi=\eta=\frac{-3 m_{\eta \Phi \Phi}\pm [9 m_{\eta \Phi
\Phi}^2-32(m_\Phi^2+m_\eta^2)(\lambda_\Phi+\lambda_\eta+4\lambda_{\eta
\Phi})]^{1/2}}{4(\lambda_\Phi+\lambda_\eta+4\lambda_{\eta
\Phi})},$$ where $+$ is for negative $m_{\eta \Phi \Phi}$ and $-$
is for positive $m_{\eta \Phi \Phi}$.
 Eq.~(\ref{split}) combined with Eq.~(\ref{bbb}) imply
$$|m_\eta-2
m_{DM}|<\lambda_4\sqrt{\lambda_\Phi+\lambda_\eta+4\lambda_{\eta
\Phi}}~{\rm GeV}.$$ This means a mild fine tuning between $m_\eta$
and $2m_{DM}$ is required to obtain $\sigma_{tot}\sim
10^{-36}~{\rm cm}^2$.

Lepton number violating terms such as $\eta \Phi^\dagger \Phi$ and
$\eta^\dagger \Phi \Phi$ can be added to the Lagrangian but since
their couplings should be much smaller than $m_{\eta \Phi \Phi}$,
they cannot change the discussion. Moreover, once the lepton
number is broken, a small mass term of form $\tilde{m}_\Phi^2
\Phi\Phi/2$ can be also added. This means that there can be a
splitting between the imaginary and real components of $\Phi$.
Notice that a real $m_{\eta \Phi \Phi}$ coupling leads to the
annihilation of a pair of real components together and a pair of
imaginary components together. This is unlike the annihilation
through a neutral gauge boson that takes place between the
imaginary and real components of $\Phi$. As a result, a small
splitting between the real and imaginary components will not
change the annihilation processes. However, the heavier component
can decay into the lighter one and a pair of neutrinos via the
mixing of $\Delta^0$ and $\eta$ and via the $f_{\alpha \beta}$
coupling. If the decay takes place when the DM particles have
become non-relativistic, the energy of neutrinos will be given by
the mass difference between lighter and heavier components of
$\Phi$ which is given by $|\tilde{m}_\Phi^2|/(2 m_{DM})$. If this
splitting is much smaller than 1~MeV (which is natural with taking
$\tilde{m}_\Phi^2/(2 m_{DM})\sim \mu\ll 1~$MeV), the energy of
these neutrinos will be too small to destroy the products of
nucleosynthesis even if the decay takes place at or after
nucleosynthesis era.

\item\textbf{Dirac fermion, $\psi$, as DM}

The general $Z_2$ invariant and lepton number conserving
Lagrangian involving $\psi$ can be written as
$$\lambda_5 \eta \frac{\psi^T_L{\mathcal{C}} \psi_L}{2}+\lambda_6 \eta
\frac{\psi^T_R{\mathcal{C}} \psi_R}{2}+ m_{DM} \bar{\psi}_R\psi_L+
{\rm H.c.}$$ Again through a $s$-channel diagram, this Lagrangian
leads to
$$\sigma({\rm DM}+{\rm DM} \to  \nu_\alpha \nu_\beta) =
\frac{(\lambda_5^2+\lambda_6^2)}{64 \pi }\left(\frac{\lambda_4
f_{\alpha \beta} m_{DM}v^2 }{[m_\Delta^2-(2 m_{DM})^2]
[m_\eta^2-(2 m_{DM})^2]}\right)^2\ .$$ To obtain $\sigma_{tot}
\sim 10^{-36}~{\rm cm}^2$, \be \left|
m_\eta^2-(2m_{DM})^2\right|\sim 300~{\rm GeV}^2~\frac{\lambda_4
m_{DM}}{500~{\rm GeV}} \frac{f_{\alpha \beta}}{0.01}
\left(\frac{1~{\rm
TeV}}{m_\Delta}\right)^2\left(\frac{\lambda_5^2+\lambda_6^2}{2}\right)^{1/2}
.\ee Similarly to the case with complex scalar DM, to obtain the
required value of $\sigma_{tot}$ a fine tuning between $m_\eta$
and $2m_{DM}$ is required.

 Lepton number violating mass terms of
form $m_R \psi_R^T{\mathcal{C}}\psi_R/2$ and $m_L
\psi_L^T{\mathcal{C}}\psi_L$ can be added to the Lagrangian. These
terms lead to a mass splitting. Remember that we discussed the
effects of decay and annihilation in the presence of mass
splitting for scalar DM. The same discussion applies here, too.

\end{itemize}

Notice that in this model the flavor structure of ${\rm DM +DM}
\to \nu_\alpha \nu_\beta$ is given by $|(m_\nu)_{\alpha
\beta}|^2$. It is also possible to have  a two component DM
scenario with both $\psi$ and $\Phi$. A $Z_2\times Z_2$ symmetry
can stabilize both of them.

Through the coupling of the DM with $\eta$ and the mixing of
$\eta$ with $h$, the DM can interact with nuclei but the
scattering cross section will be suppressed by the lepton number
violating parameter $\mu^\prime$; {\it i.e.,} by the mixing of $h$
and $\eta$.
 To obtain significant scattering cross section
off nuclei, we add a real scalar $\xi$ as the portal to quark
sector through mixing with the SM Higgs. The term that mixes $H$
and $\xi$ is
$$m_{\xi H H} \xi H^\dagger H$$
which leads to a mixing of
$$ \tan 2\alpha_{h\xi}=\frac{2m_{\xi HH} v}{m_\xi^2-m_h^2}\ . $$

The DM candidates also couple to $\xi$. The coupling to the scalar
DM, $\Phi$,
$$ m_{\xi \Phi\Phi} \xi \Phi^\dagger \Phi \ .$$
leads to
$$\sigma({\rm DM}+ N \to {\rm DM}+  N)=\sigma(\Phi N \to \Phi N) =\frac{f_N^2}{\pi}\left(
\frac{m_{\xi\Phi\Phi}m_{\xi HH}}{m_\xi^2}\right)^2 \frac{\mu_{{\rm
DM}N}^2 m_N^2}{m_{DM}^2 m_h^4}$$
$$\sim 10^{-8} ~  pb\left(\frac{m_{\xi\Phi\Phi}m_{\xi
HH}/m_\xi^2}{0.1}\right)^2\left( \frac{200~{\rm
GeV}}{m_{DM}}\right)^2\left( \frac{120~{\rm GeV}}{m_h}\right)^4
\left(\frac{f_N}{0.3}\right)^2,$$ where $N$ collectively denotes
nucleons ({\it i.e.,} $n$ and $p$) and $\mu_{{\rm DM} N}\simeq
1$~GeV is the reduced mass of the $\Phi$-$N$ system \cite{Hambye}.

The same coupling and mixing lead to $\Phi+\Phi^\dagger \to
\xi^*,h^* \to f+\bar{f}, W^++W^-,{Z}+Z$ with $\langle \sigma
v\rangle$ equal to \be \label{sub} (2m_{DM}\Gamma(h\to {\rm
final~states}))|_{m_h\to 2m_{DM}} \frac{m_{\xi \Phi\Phi}^2 m_{\xi
HH}^2v^2}{4 m_{DM}^2(4m_{DM}^2-m_\xi^2)^2(4m_{DM}^2-m_h^2)^2}.\ee
$\Gamma(h\to {\rm final~states}))$ versus $m_h$ can be found in
\cite{Hdecay}. Taking $m_{\xi \Phi \Phi}m_{\xi
HH}/|4m_{D}^2-m_\xi^2|\stackrel{<}{\sim} 0.1$ and
$m_{DM}<300$~GeV, $\langle \sigma_{tot} v \rangle$ will be smaller
than $10^{-36}~{\rm cm}^2$. However, for $m_{DM}\sim 200-400$~GeV,
the annihilation mode via $s$-channel $h$ and $\xi$ exchange can
be  significant  along with $\Phi+\Phi \to \nu+\nu$. As discussed
in \cite{Esmaili2}, in this case still the seasonal variation of
muon track events at neutrino telescopes can be significant.

 Similarly, the
coupling to the fermionic DM, $\psi$,
$$ Y_{\xi \psi} \xi \bar{\psi} \psi+ H.c.$$
results in \be \label{refer}\sigma({\rm DM}+ N \to {\rm DM}+  N)
=\sigma(\psi N \to \psi N) =2\frac{f_N^2}{\pi}\left( \frac{ Y_{\xi
\psi}  m_{\xi HH}}{m_\xi^2}\right)^2 \frac{\mu_{{\rm DM} N}^2
m_N^2}{m_h^4}\sim \ee
$$
10^{-8} ~  pb\left(\frac{ Y_{\xi \psi} m_{\xi HH}\times 3~{\rm
TeV}}{m_\xi^2}\right)^2\left( \frac{120~{\rm GeV}}{m_h}\right)^4
\left(\frac{f_N}{0.3}\right)^2\ $$ where $\mu_{{\rm DM} N}\simeq
1$~GeV is the reduced mass of the $\psi$-$N$ system \cite{Hambye}.
In both cases, $0.14<f_n,f_p<0.66$. Thus, within the favored range
of parameters ($m_{DM}\sim 200$ GeV), the scattering cross section
will be of order of $10^{-8}~pb$ which is high enough to lead to a
few hundred neutrino events (or even more for larger $m_{DM}$) at
ICECUBE each year. We can have a detectably large monochromatic
neutrino flux from DM annihilation inside the Sun with a
non-democratic flavor composition determined by
$(m_\nu)_{\alpha\beta}$. The $ Y_{\xi \psi}$ coupling also leads
to $\psi+\bar{\psi}\to f+\bar{f}, W^++W^-,{Z}+Z$ with
 $\langle \sigma
v\rangle$ equal to \be \label{subsub} (2m_{DM}\Gamma(h\to {\rm
final~states}))|_{m_h\to 2m_{DM}} \frac{Y_{\xi \psi}^2 m_{\xi
HH}^2v^2/2}{(4m_{DM}^2-m_\xi^2)^2(4m_{DM}^2-m_h^2)^2}v_{rel}^2\ee
 where $v_{rel}$ is the relative velocity of the DM
pair. Notice that in Eq.~(\ref{sub}) for the scalar DM case, such
a factor of $v_{rel}^2$ does not appear. At the decoupling era,
$v_{rel}\sim 1/\sqrt{20}$ so for $\sqrt{2}m_{DM}m_{\xi HH} Y_{\xi
\psi}/|4m_{DM}^2-m_\xi^2|\stackrel{<}{\sim} 0.1$, the annihilation
to the Higgs decay products in the early universe can be only
subdominant: $\langle \sigma(\psi +\bar{\psi}\to
 \xi^*,h^*\to {\rm anything})v\rangle/\langle \sigma_{tot} v \rangle<0.01$. For the DM
particles trapped inside the sun,
$$v_{rel}^2\sim \frac{3 k_BT}{m_{DM}}\sim 10^{-9} \frac{300~{\rm
GeV}}{m_{DM}}\frac{k_BT}{100~{\rm eV}}$$ so the annihilation to
the Higgs decay products can be safely neglected. Thus, for the
purpose of this paper, the $v_{rel}^2$ dependence is favored. Had
we taken the  interaction of $\xi$ with $\psi$ to be of form $i
\xi \bar{\psi} \gamma^5 \psi$ instead of $\xi \bar{\psi} \psi$,
such factor of  $v_{rel}^2$ would not have appeared in
Eq.~(\ref{subsub}).

Let us now discuss the flavor structure of the neutrinos produced
by the DM annihilation. The amplitude of ${\rm DM}+{\rm DM}\to
\nu_\alpha+\nu_\beta$, ${\mathcal{M}}_{\alpha\beta}$, is
proportional to $f_{\alpha\beta}$ which is in turn proportional to
$(m_\nu)_{\alpha\beta}$. This means in the mass basis where
$(m_\nu)$ is diagonal, the amplitude ${\mathcal{M}}$ is also
diagonal. Let us denote the neutrino mass eigenstates in vacuum by
$|i\rangle=|1\rangle,|2\rangle$ and $|3\rangle$. The neutrino
production is in the form $|1\bar{1}\rangle$, $|2\bar{2}\rangle$
and $|3\bar{3}\rangle$ (but not for example of form
$|1\bar{2}\rangle$). The production rate of $|i\bar{i}\rangle$ is
given by $|f_{ii}|^2 \propto |m_i|^2$. That is the density matrix
in the mass basis is proportional to Diag$(m_1^2,m_2^2,m_3^2)$. As
a result, for the quasi-degenerate neutrino mass scheme with
$|m_1|^2\simeq |m_2|^2\simeq |m_3|^2$, the neutrino production
will be democratic and as discussed in detail in Ref.
\cite{Esmaili2}, the seasonal variation will vanish. Let us now
evaluate the seasonal variation for the general neutrino mass
scheme. Due to the matter effects inside the Sun, a pure mass
eigenstate while crossing the Sun converts to a combination of
mass eigenstates as follows: $|i;{\rm surface}\rangle=\sum_j
a_{ij}|j\rangle$ and $|\bar{i};{\rm surface}\rangle=\sum_j
\bar{a}_{ij}|\bar{j}\rangle$ where due to the matter effects
$a_{ij}$ may differ from $\bar{a}_{ij}$. $a_{ij}$ and
$\bar{a}_{ij}$ are unitary matrices: $\sum_j
\bar{a}_{ij}\bar{a}_{kj}^*=\delta_{ik}$ and  $\sum_j
{a}_{ij}{a}_{kj}^*=\delta_{ik}$. After traversing the distance
between the Sun and Earth, $L$, $|j\rangle$ and $|\bar{j}\rangle$
will pick up a phase of $e^{-i m_j^2 L/2E}$. Neglecting the
average of the phase $e^{i\Delta m_{31}^2 L/2E}$, we can write
$\langle P(\nu_i \to \nu_\mu)\rangle=\sum_j |U_{\mu j}|^2 |a_{i
j}|^2+2\Re [ a_{i1}^* a_{i2} U_{\mu 1} U_{\mu 2}^* e^{i \Delta
m_{21}^2 L/2E}]$ (see Ref. \cite{Esmaili2}). Considering the fact
that the production of $|i\rangle$ is given by $m_i^2$, the
seasonal variation due to $\Delta m_{12}^2L/2E\sim 1$ can be
evaluated as
$${\rm variation}=\frac{2 \sum_i m_i^2 \Re [ a_{i1}^* a_{i2}
U_{\mu 1} U_{\mu 2}^*]}{\sum_{ij} m_i^2 |a_{ij}|^2|U_{\mu
j}|^2}.$$ Since $\Delta m_{21}^2\ll\Delta m_{31}^2$, we can take
$m_1^2\simeq m_2^2$ and can therefore write \be \label{var}{\rm
variation}=\frac{2 \sum_i \Delta m_{31}^2 \Re [ a_{31}^* a_{32}
U_{\mu 1} U_{\mu 2}^*]}{m_1^2+\Delta m_{31}^2\sum_{ij}
|a_{3j}|^2|U_{\mu j}|^2}\ee where we have used the unitarity of
$a_{ij}$. If $\theta_{13}$ is exactly zero, despite the matter
effects, $|3\rangle$ will not change which means
$a_{31}=a_{32}=0$, so the variation will vanish. Remembering that
the matter density of the Sun falls as $e^{-r/ (0.1 R_{{\rm
sun}})}$ we find that for $\sin \theta_{13}\stackrel{>}{\sim}
[(0.1 R_{{\rm sun}}) V_e|_{center}]^{-1} \sim 0.001$, the values
of $a_{31}$ and $a_{32}$ can  in general be of order of 1 which
means the variation can be sizeable (larger than 10 \%) and can be
measured by a few hundred events. Eq.~(\ref{var}) also confirms
that for quasi-degenerate  mass scheme with $m_1^2\gg |\Delta
m_{31}^2|$, the variation is suppressed even for
$a_{31}^*a_{32}\sim 1/2$.

 Let us now briefly discuss the impact of this
model for the collider searches. Like the case of type II seesaw
mechanism, we expect a triplet scalar which is produced by
electroweak interactions. Like the type II seesaw mechanism, the
$\Delta^+$ and $\Delta^{++}$ components decay into lepton pairs
via the $f_{\alpha \beta}$ coupling. For a review of the possible
decay modes of $\Delta$, see \cite{DeltaDECAY}. However, there
will be a new decay mode for $\Delta^0$, which unlike $\Delta^0
\to \nu_\alpha \nu_\beta$, is visible. The new decay mode is
$\Delta^0\to \eta h$. If $\eta$ is heavier than $2m_{DM}$, it can
decay into a DM pair which appears as missing energy. Since in
this model $H$ mixes with $\xi$, the signature of Higgs can be in
principle different. In fact, any process studied for the SM ({\it
e.g.,} $gg \to h$ and $h\to \tau \bar{\tau}, \ \gamma\gamma$)
should be reconsidered for two SM-like Higgs fields ({\it i.e.,}
two mass eigenstates composed of $h$ and $\xi$) with rates
suppressed by $\cos^2 \alpha_{\eta \xi}$ and $\sin^2 \alpha_{\eta
\xi}$. The production of the second Higgs-like scalar will be
suppressed by a factor of $\sin^2 \alpha_{\eta \xi}$.  For $\sin
\alpha_{h \xi}\stackrel{<}{\sim}0.1$, which is the favored range
by bounds on $\sigma({\rm DM}+N\to {\rm DM}+N)$,  the production
of the second Higgs can be neglected. In this case, the Higgs
sector will be similar to what we had within the Standard Model.
%%%%%%%%%%%%%%%%%
\section{MODEL II \label{modelII}}
%%%%%%%%%%%%%%%%%%%%%

In this section, we introduce another model within which DM is
composed of Dirac fields, $\psi$, which are protected against
decay again with a $Z_2$ symmetry. We introduce a new
$U(1)^\prime$ symmetry with gauge boson $Z^\prime$ under which all
the SM particles are invariant. In this model, $\nu_S$ with masses
close to those of active neutrinos also exist such that
oscillation can take place between the sterile and active
neutrinos.

The fields that are added to this model are the following:
\begin{itemize}
\item A Dirac field, $\psi=(\psi_L \ \psi_R)$, which plays the role of the DM
candidate;
\item One (or more) left-handed sterile neutrino, $\nu_S$;
\item Right-handed neutrinos $\nu_{R\beta}$ which are singlets  under both electroweak symmetry and
$U(1)^\prime$. These neutrinos are included to give Dirac masses
to the rest of neutrinos.
\item $U(1)^\prime$ gauge boson, $Z^\prime$;
\item A complex scalar field $H^\prime$ which is an electroweak
singlet but under $U(1)^\prime$ has a charge equal to that of
$\psi_L$. $H^\prime$ receives a Vacuum Expectation Value  (VEV) of
$v'/\sqrt{2}$ which breaks  the $U(1)^\prime$ symmetry and gives a
mass of $m_{Z^\prime}=g^\prime v^\prime/\sqrt{2}$ to $Z^\prime$.
After electroweak and $U(1)^\prime$ symmetry breaking, the
$H^\prime$ also mixes with $H$ and acts as a messenger  to
interact with nuclei.
\end{itemize}

Via a $s$-channel $Z^\prime$ exchange, the DM particles annihilate
into sterile neutrinos which in turn oscillate into active
neutrinos with a non-trivial flavor composition given by the
flavor structure of the mixing of the sterile neutrinos with the
active neutrinos.

We assume that the right-handed $\psi$ is neutral under
$U(1)^\prime$ and only left-handed $\psi$ couples to $Z^\prime$:
$${\mathcal{L}}=g'\left( \bar{\nu}_S \gamma^\mu
 (\frac{1-\gamma^5}{2})\nu_S-\bar{\psi} \gamma^\mu
 (\frac{1-\gamma^5}{2})\psi\right) Z_\mu'\ .$$
 With this charge assignment there is no $U(1)^\prime$ anomaly.
If we assigned the same $U(1)^\prime$ charge to $\psi_R$, we must
add additional chiral fermions to cancel the $U(1)^\prime$
anomaly.

Through this coupling, the DM pair annihilates into a sterile
neutrino pair with the annihilation cross section
$$ \langle\sigma( \bar{\psi}\psi \to \bar{\nu}_S \nu_S)\rangle = \frac{
g'^4}{8 \pi}\frac{m_{DM}^2}{\left[(2m_{DM})^2-m_{Z'}^2\right]^2}$$
where the mass of $\nu_S$ which is taken to be of order of the
active neutrino masses is neglected ({\it i.e.}, $m_{\nu_S}\ll
m_{DM}$). To have $\langle \sigma_{tot} v\rangle=10^{-36}~{\rm
cm}^2$ ({\it i.e.,} the value suggested by the DM abundance in
thermal DM production scenario), we find
$$
m_{Z^\prime}^2=4m_{DM}^2-\left(\frac{g^\prime}{0.16}\right)^2
m_{DM}\times (100~{\rm GeV}) \ .$$ For example, for
$g^\prime=0.16$ and $m_{DM}=200~$GeV, the desired abundance of DM
can be achieved with $m_{Z^\prime}=375$~GeV. The $Z^\prime$ boson
in this model does not couple to quarks or leptons and  there is
no mixing between $Z^\prime$ and the SM gauge bosons. As a result,
none of the bounds from the new gauge boson searches at colliders
applies in this case.

%In this model DM pair which is a Dirac field annihilate to a pair
%sterile neutrino which in turn oscillates to active neutrinos.
%Both DM and sterile neutrinos are charged under a new $U(1)'$
%gauge with gauge boson $Z'$. To give mass to $Z'$ a new Higgs $H'$
%is added which is singlet under SM gauge but charged under new
%$U(1)'$. Through mixing with SM Higgs this field acts as a portal
%with quarks. There are also total singlet right-handed neutrinos
%that help to give Dirac mass to  both active and sterile neutrinos
%and mix them.
% Gauge couplings:

 After breaking of $U(1)'$  and the electroweak symmetry, the neutrinos and $\psi$ receive Dirac  mass
 which comes from the following Yukawa couplings:
$${\mathcal{L}}= f_{\alpha \beta} \bar{\nu}_{R\beta} L_\alpha^i
H^j \epsilon_{ij}+ f_{R \beta} \bar{\nu}_{R\beta} H'\nu_{S}+
Y_\psi \bar{\psi}_R\psi_L H'^\dagger + {\rm H.c.}\ .$$ We assumed
that only $\psi$ is odd under the $Z_2$ symmetry so the $Z_2$
symmetry prevents a coupling of form $\bar{\nu}_{R\beta}\psi_L
(H^\prime)^\dagger$ or a mass term of form
$\nu_{R\beta}^Tc\psi_R$. The masses of $\psi_L$ and $\psi_R$ are
the same and equal to
$$m_{DM}=Y_\psi\langle H^\prime\rangle =Y_\psi
\frac{v^\prime}{\sqrt{2}}.$$ After $H$ and $H^\prime$ obtain
vacuum expectation values, $\nu_S$ receives a Dirac mass and mixes
with active neutrinos. For illustrative purposes, let us consider
one sterile neutrino, one active flavor and two right-handed
neutrinos. The masses will be given by \ba \frac{1}{\sqrt{2}}[
\bar{\nu}_{R1} \ \bar{\nu}_{R2}]\left[
\begin{matrix} f_{1 \alpha}v & f_{R 1} v' \cr f_{2 \alpha}v & f_{R 2} v'
\end{matrix} \right] \left[ \begin{matrix}  \nu_\alpha \cr \nu_S
\end{matrix} \right]\label{ASmixing}
\ea The mixing between sterile and active  neutrinos is given by
the following formula $$\tan 2 \theta=\frac{2
(f_{1\alpha}f_{R1}+f_{2\alpha}f_{R2}) v v'
}{(f_{1\alpha}^2+f_{2\alpha}^2)v^2-(f_{R1}^2+f_{R2}^2)v'^2}$$

 The
Lagrangian of the model includes a term as follows
$$ Y_{h'h} H'^\dagger \cdot H' H^\dagger \cdot H.$$
In the unitary gauge $H=(0 \ (v+h)/\sqrt{2})$ and $H^\prime=
(v^\prime +h^\prime)/\sqrt{2}$:  \ba \frac{1}{{2}}\ [ h \ h']
\left[\begin{matrix} m_h^2 & Y_{h'h} v v'\cr
 Y_{h'h} v v' &  m_{h'}^2 \end{matrix}\right] \left[\begin{matrix} h \cr
 h'\end{matrix}\right] \ . \ea
 The mixing between $h$ and $h'$ is given by $$\tan 2 \alpha_{hh'} =
 \frac{2Y_{h'h} v v' }{ m_h^2-m_{h'}^2}\ .$$
 This mixing leads
to DM scattering off the nuclei in the target  of the direct DM
search experiments with a cross section
$$\sigma({\rm DM}+ N \to {\rm DM}+ N) =\frac{f_N^2}{\pi}\left(
\frac{Y_\psi Y_{h'h} v^\prime}{m_{h'}^2m_h^2}\right)^2 \mu_{{\rm
DM} N}^2 m_N^2=$$ $$ 10^{-8} p{\rm b}\times \left(\frac{Y_\psi
Y_{h'h} v^\prime/m_{h'}}{0.1}\right)^2\left( \frac{200~{\rm
GeV}}{m_{h'}}\right)^2\left( \frac{120~{\rm GeV}}{m_h}\right)^4
\left(\frac{f_N}{0.3}\right)^2\ .$$
 As expected, this cross section is the same as the one in Eq.~(\ref{refer}) provided that
we replace $m_\xi \to m_{h'}$, $m_{\xi HH}\to Y_{h'h} v'$ and $
Y_{\xi \psi} \to Y_\psi/\sqrt{2}$.
 As mentioned before, $0.14<f_n,f_p<0.66$. The cross section is
large enough to lead to a significant capture rate and
subsequently to a large neutrino flux detectable at ICECUBE. Again
 replacing  $m_\xi \to m_{h'}$, $m_{\xi HH}\to Y_{h'h} v'$ and $
Y_{\xi \psi} \to Y_\psi/\sqrt{2}$ in Eq.~(\ref{subsub}), the
formula for $\sigma({\rm DM}+{\rm DM} \to \xi^*,h^*\to
W^-W^+,Z{Z},t\bar{t})$ can be obtained. Similarly to the case of
Eq.~(\ref{subsub}), we find that for $Y_{hh'}  Y_{ \psi} m_{DM}
v'/|4m_{DM}^2-m_{h'}^2|<0.1$, this annihilation mode  can be
safely neglected.

Among the new particles that are added to this model, only $h'$
has a significant coupling to the SM. As a result, only $h'$ might
appear in the collider searches through mixing with the Higgs
field. The same discussion in the case of Model I described in the
previous section applies here, too: If $\sin \alpha_{h h'}<0.1$,
the production of the new Higgs at collider will be suppressed by
$\sin^2 \alpha_{h h'}<0.01$ relative to the  ordinary SM Higgs.

 Let us
now address the oscillation of sterile neutrinos to active
neutrinos and the flavor composition of the neutrino flux from DM
annihilation when they reach the detector.  This problem has
recently been addressed in \cite{Kopp}. In Eq.~(\ref{ASmixing}),
for simplicity we have assumed that $\nu_S$ mixes only with one
active flavor. In general, $\nu_S$ can mix with more than one
flavor. In fact, explaining the LSND and MiniBooNE results in the
context of the present scenario suggests that the sterile neutrino
simultaneously mixes with $\nu_e$ and $\nu_\mu$. Recent fits to
the short baseline neutrino data can be found in \cite{fit}. Some
hints for the oscillation of atmospheric $\nu_\mu$ to $\nu_S$ has
also been found in the ICECUBE data \cite{vernon}. The scenario
proposed in this section can include two sterile neutrinos mixing
with different flavors as required by the $3+2$ scenario suitable
for explaining the LSND and MiniBooNE results. Some hints for
sterile neutrinos have also been found by studying the solar
neutrino oscillation data \cite{Holanda}. Studying the general
case is beyond the scope of the present paper. We focus on
the particular case that %$\nu_S$ mixes only with $\nu_\tau$. That is we take
the mixing matrix in the $(\nu_e \ \nu_\mu \ \nu_\tau \ \nu_S)$
basis is of the following form
$$U^{(4)}=O_{34}\cdot (U_{PMNS} \oplus 1_{1\times 1})$$ where \ba
O_{34}=\left[
\begin{matrix} 1 & 0 & 0 & 0 \cr 0 & 1 & 0 & 0 \cr  0 & 0 & \cos
\theta_{34} & \sin \theta_{34}\cr  0 & 0 & -\sin \theta_{34} &
\cos \theta_{34}
\end{matrix}\right] \ , \ea
$ (U_{PMNS} \oplus 1_{1\times 1}$) is the external sum of the
standard three by three mixing matrix of neutrinos $U_{PMNS}$ with
$1_{1\times 1}$ which means its first $3\times 3$ block is
$U_{PMNS}$, its 44 element is equal to one and the rest of its
elements vanish. Notice that, without matter effects,
P($\stackrel{(-)}{\nu_e} \to \stackrel{(-)}{\nu_e}$),
P($\stackrel{(-)}{\nu_\mu} \to \stackrel{(-)}{\nu_\mu}$) and
P($\stackrel{(-)}{\nu_\mu} \to \stackrel{(-)}{\nu_e}$) as well as
P($\stackrel{(-)}{\nu_e }\to \stackrel{(-)}{\nu_\mu}$) are the
same as what we expected without the sterile neutrinos. As a
result, the bounds from reactor searches as well as the
measurement of P($\stackrel{(-)}{\nu_\mu }\to
\stackrel{(-)}{\nu_e}$) and P($\stackrel{(-)}{\nu_\mu} \to
\stackrel{(-)}{\nu_\mu}$) by atmospheric and long baseline
experiment and various other experiments do not apply here. This
mixing  affects P($\stackrel{(-)}{\nu_\mu} \to
\stackrel{(-)}{\nu_\tau}$) which is poorly constrained by
observation. The strongest constraint comes from the measurement
of the total neutral current interaction of the beam in the MINOS
experiment \cite{minos}. If this constraint is saturated,
$\theta_{34}$ can be still relatively large leading to a
significant probability of sterile to active oscillation.
%As discussed in sect. \ref{review}, in this scenario,
%the monochromatic $\nu_S$ flux (even for
%$E_{\nu_S}=m_{DM}>500$~GeV) will leave the Sun without being
%affected by scattering off the nuclei. That is the monochromatic
%flux will remain monochromatic.

 A part of the $\nu_S$ flux will
oscillate into active neutrinos on the way to the Earth which can
be detected  at ICECUBE or its deepcore (provided that the
detection threshold is lowered below $m_{DM}$). The number of muon
track events in a time interval is given by $\langle P(\nu_S\to
\nu_\mu )\rangle$ which is the oscillation probability  averaged
over the time interval. As discussed in \cite{Esmaili2}, due to
the averaging effects the oscillatory terms given by $\Delta
m_{31}^2$ are subdominant however we should keep the oscillatory
terms given by $\Delta m_{21}^2$. A simple numerical calculation
shows that the matter effects on the oscillation probability
cannot be neglected. Adopting the formalism in \cite{Esmaili2} and
generalizing it to  a four-neutrino scheme, we can expand the
evolved $\nu_S$ at the Sun surface in terms of the mass
eigenstates as follows
$$|\nu_S;{\rm surface}\rangle = a_{S1}|1 \rangle +a_{S2}|2 \rangle
+a_{S3}|3 \rangle +a_{S4}|4 \rangle \ .$$ By a numerical
calculation taking into account matter effects,  $a_{Si}$ can be
found in terms of the $U^{(4)}$ elements and $\Delta m_{34}^2$.
Averaging out the oscillatory terms given by $\Delta m_{31}^2$, we
can then write
$$\langle  P(\nu_S\to \nu_\mu)\rangle=|U_{\mu 1}|^2|a_{S1}|^2+|U_{\mu 2}|^2|a_{S2}|^2
+|U_{\mu 3}|^2|a_{S3}|^2+$$ $$ 2\Re[U_{\mu 1}U_{\mu
2}^*a_{S1}^*a_{S2}\exp (i\frac{\Delta m_{21}^2 L_{{\rm
Sun-Earth}}}{2 E_\nu})]\ ,$$ where $E_\nu=m_{DM}$.
 As
discussed in \cite{Esmaili1}, by studying the periodicity of muon
track events, the value of $\Delta m_{12}^2/m_{DM}$ can be
derived. The results of a simple numerical calculation shows that
if the bound from \cite{minos} on $\theta_{34}$ is saturated,
P$(\nu_S\to \nu_\mu)$ can reach as large as 0.2. This means for
$\sigma({\rm DM}+N\to {\rm DM}+N)\sim 10^{-8}~{\rm cm}^2$, up to a
few hundred muon-track events can be registered each year at
ICECUBE so the statistics will be enough to study a seasonal
variation \cite{Esmaili2}.
 In ref. \cite{Esmaili2}, we have found that for three-neutrino scheme
 in the case that ${\rm DM}+{\rm DM}\to \nu_\mu+\nu_\mu$ or
 $\nu_\tau+\nu_\tau$, the variation will be small. Obviously this
 consideration does not apply to the present model with four
 neutrinos and ${\rm DM}+{\rm DM}\to \nu_S+\nu_S$.

 Notice that $\langle P(\nu_S\to
\nu_\mu)\rangle$ is not sensitive to $\Delta m_{34}^2$ because we
have taken $U_{\mu 4}=0$. The total oscillation probability to
active neutrinos, $\langle 1- P(\nu_S\to \nu_S)\rangle$, is given
by
$$|U_{S 1}|^2|a_{S1}|^2+|U_{S 2}|^2|a_{S2}|^2 +|U_{S
3}|^2|a_{S3}|^2+|U_{S 4}|^2|a_{S4}|^2+$$
$$ 2\Re[U_{S 1}U_{S 2}^*a_{S1}^*a_{S2}\exp (i\frac{\Delta
m_{21}^2 L_{{\rm Sun-Earth}}}{2 E_{\nu}})]+$$ $$2\Re[U_{S 3}U_{S4
}^*a_{S3}^*a_{S4}\exp (i\frac{\Delta m_{43}^2 L_{{\rm
Sun-Earth}}}{2 E_{\nu}})]\ .$$The number of the cascade-like
events is given by $\sigma_{NC}\langle 1- P(\nu_S\to
\nu_S)\rangle+\sigma_{CC}\langle 1- P(\nu_S\to \nu_S)-P(\nu_S\to
\nu_\mu)\rangle$. Thus, by studying the variation of cascade-like
events, it will be in principle possible to derive $\Delta
m_{34}^2$. The possibility  can be realized only if the following
three conditions are fulfilled: (i) The energy threshold for
cascade detection is below $m_{DM}$; (ii) The statistics is high
enough; (iii) The  oscillation length $L_{osc}=4 \pi m_{DM}/\Delta
m_{34}^2$ is of the order of the seasonal variation of the
Sun-earth distance: {\it i.e.,} $L_{osc}\sim\Delta L_{{\rm
Sun-Earth}}\sim 5~{\rm million}~ {\rm km}$, which means
$$\Delta m_{34}^2 \sim 10^{-5}~{\rm
eV}^2\left(\frac{m_{DM}}{200~{\rm GeV}}\right).$$ It is noteworthy
that for $m_{DM}\sim 200$~GeV, this means $\Delta m_{12}^2\sim
\Delta m_{34}^2$. Unlike the case of $\mu$-track events, the
directionality cannot be used to reduce the background of
shower-like events from atmospheric neutrinos. At ICECUBE, we
expect a maximum of a few hundred cascade-like neutrino events
from the DM annihilation inside the Sun per year which is within
the statistical fluctuation of the background from the atmospheric
neutrinos. This means that after 10 years of data taking by
ICECUBE, the confidence level of the discovery of such
cascade-like signal will be about 3-4 $\sigma$.
 To make a serious measurement, a detector ten times larger than
 ICECUBE with threshold below $m_{DM}$
 is required. After about 2-3 years of data taking by such a
 detector, the statistics can be enough to claim discovery and
 after ten years, enough data can be collected to study variation
 and to extract information from the variation.

% In this scenario, the ratio of the contained
%muon-track events to shower-like events is determined by
%$$\frac{\sigma_{CC} {\rm Br}(\tau\to \mu
%\nu\nu)}{\sigma_{NC}+(1-{\rm Br}(\tau\to \mu
%\nu\nu))\sigma_{CC}}.$$

 Let us now discuss the possibility of
constraining this scenario by various other oscillation
observations. As discussed above, P$(\bar{\nu}_e\to \bar{\nu}_e)$
relevant for the reactor neutrino data is not affected by the
presence of the sterile neutrino. However, because of the matter
effects inside the Sun, the survival probability of the solar
$\nu_e$ will be affected by the presence of the sterile
 neutrino. The effect in general is rather complicated. For our
 case with $|\Delta m_{41}^2|\simeq |\Delta m_{31}^2|\gg \Delta
 m_{21}^2$, the simplified formalism in \cite{Giunti} can be
 applied. Using the formalism in \cite{Giunti}, we find that the
 deviation from SM neutrino oscillation without sterile neutrino
 is suppressed by $(N_n/2N_e)\cos 2\theta_{12}c_{23}^2 s_{34}^2$
 where $N_n$ and $N_e$ are respectively the neutron and electron
 number densities in the Sun. Inserting the numerical values, we
 find that the deviation is of order of $0.02 \sin^2 \theta_{34}$
 and is therefore very small even for $\sin^2 \theta_{34}\sim 1$.

 Measuring P($\stackrel{(-)}{\nu_\mu} \to
\stackrel{(-)}{\nu_\tau}$) at long baseline experiment OPERA or by
studying the atmospheric neutrino data at Super-Kamiokande
constrains $\theta_{34}$ as in this scenario
P($\stackrel{(-)}{\nu_\mu} \to \stackrel{(-)}{\nu_\tau}$) is
suppressed by a factor of $\cos^2 \theta_{34}$ relative to the
standard three-neutrino scheme. This scenario can be also probed
by improving the measurement of the rate of neutral current
interaction events at long baseline experiments such as MINOS.

\section{Summary \label{con}}
We have introduced two models within which the DM pair dominantly
annihilates into a neutrino pair with a non-trivial flavor
composition. As a result, the non-relativistic DM pair
annihilation will lead to a sharp line in the neutrino energy
spectrum with $ F_{\stackrel{(-)}{\nu_e}}:
F_{\stackrel{(-)}{\nu_\mu}}:F_{\stackrel{(-)}{\nu_\tau}}\ne
1:1:1$. As shown in \cite{Esmaili1,Esmaili2}, despite the very
large distance between the Sun and Earth, the oscillation
probability does not average to zero, leading to a seasonal
variation as the distance between the Sun and Earth varies during
a year. The variation contains information on $m_{DM}$ and the
initial flavor composition.

In these models, the DM particles are stabilized by a $Z_2$
symmetry under which only the DM candidates are odd. In both
models, the DM particles interact with nuclei via a new scalar
which mixes with the SM Higgs so the interaction is
spin-independent. The parameters are chosen to  yield a scattering
cross section in the range $10^{-9}-10^{-8}~p$b. Thus, while the
present bounds from direct searches are satisfied, still
significant statistics (a few hundred events per year) are
expected at ICECUBE for indirect DM detection. This means by a
slight improvement in both direct and indirect DM searches, these
models can be tested. The same coupling and mixing can also lead
to ${\rm DM}+{\rm DM} \to f +\bar{f},Z +{Z},W^++W^- $. However,
these annihilation modes will be subdominant. When DM is composed
of Dirac fermions, these modes, being  P-wave effects, are further
suppressed by $v_{rel}^2$ and  for the case of the DM pair trapped
inside the Sun can be safely neglected.

Model I embeds type II seesaw mechanism so neutrinos are Majorana
particles. Within this model, the DM candidates can be either
complex scalars or Dirac fermions with lepton number equal to 1.
It is also possible that both  the complex scalars and the Dirac
fermions contribute to the DM in the universe. The DM pair
annihilates into $\nu_\alpha \nu_\beta$ with a flavor composition
determined by $(m_\nu)_{\alpha \beta}$. The prediction of the
model for LFV rare decays  as well as  the accelerator searches is
similar to the predictions of type II seesaw mechanism, except
that here the neutral component of the triplet, $\Delta^0$, can
have a new decay mode, $\Delta^0 \to h +{\rm missing~energy}$.

In Model II, the DM is composed of Dirac fermions which via the
exchange of a new $Z^\prime$ gauge boson annihilate into a pair of
sterile neutrinos. Since $Z^\prime$ does not couple to quarks or
ordinary leptons, it cannot be produced in the lepton or hadron
colliders. In this model, neutrinos are Dirac particles so we
expect a null result in searches for neutrinoless double beta
decay.

Since the annihilation products in the Sun center are sterile,
they do not scatter off the nuclei present inside the Sun so, to
the first approximation, the height of the sharp line in the
spectrum is not reduced by scattering.
 On their way to Earth, the sterile neutrinos  oscillate into
active neutrinos. The flavor composition of the flux on Earth is
given by the mixing parameters of the sterile neutrinos with the
active neutrinos. The number of the muon-track events is given by
P($\nu_S\to \nu_\mu$) which in turn is given by the active-sterile
mixing. As discussed in the text, the main constraint on the
mixing comes from the MINOS measurement  of the total neutral
current interaction of the beam at the detector. Taking into
account this bound, the statistics can be still high enough to
employ the method introduced in \cite{Esmaili2}. By improving this
bound, the model will be further constrained. While the  variation
of the muon-track events  at neutrino telescope will be sensitive
to $\Delta m_{12}^2/m_{DM}$, the variation of the cascade events
at the neutrino telescopes will be sensitive to both $\Delta
m_{34}^2/m_{DM}$ and $\Delta m_{12}^2/m_{DM}$.

\section*{Acknowledgement}
The author would like to thank A. Esmaili for useful  discussions.
She is also grateful to E. Ma and T. Schwetz for fruitful
comments. She would like to acknowledge ICTP (especially the high
energy group and associate scheme) where the idea for this work
was formed.


\begin{thebibliography}{99}
\bibitem{Manfred}
M.~Lindner, A.~Merle, V.~Niro,
  %``Enhancing Dark Matter Annihilation into Neutrinos,''
  Phys.\ Rev.\  {\bf D82 } (2010)  123529.
  [arXiv:1005.3116 [hep-ph]].
\bibitem{Esmaili1}
  A.~Esmaili, Y.~Farzan,
  %``On the Oscillation of Neutrinos Produced by the Annihilation of Dark Matter inside the Sun,''
  Phys.\ Rev.\  {\bf D81 } (2010)  113010.
  [arXiv:0912.4033 [hep-ph]].
\bibitem{Esmaili2}
  A.~Esmaili, Y.~Farzan,
  %``A Novel Method to Extract Dark Matter Parameters from Neutrino Telescope Data,''
  JCAP {\bf 1104 } (2011)  007.
  [arXiv:1011.0500 [hep-ph]].

\bibitem{direct}
{\it See for example,} E.~Aprile {\it et al.} [XENON100
Collaboration ],
  %``Dark Matter Results from 100 Live Days of XENON100 Data,''
  Phys.\ Rev.\ Lett.\  {\bf 107 } (2011)  131302.
  [arXiv:1104.2549 [astro-ph.CO]].
  \bibitem{example}
J.~March-Russell, C.~McCabe, M.~McCullough,
  %``Neutrino-Flavoured Sneutrino Dark Matter,''
  JHEP {\bf 1003 } (2010)  108;
  [arXiv:0911.4489 [hep-ph]];
  K.~M.~Belotsky, M.~Y.~Khlopov, K.~I.~Shibaev,
  %``Monochromatic neutrinos from massive fourth generation neutrino annihilation in the sun and earth,''
  Part.\ Nucl.\ Lett.\  {\bf 108 } (2001)  5-17;
  K.~M.~Belotsky, M.~Y.~Khlopov, K.~I.~Shibaev,
  %``Monochromatic neutrinos from the annihilation of fourth-generation massive stable neutrinos in the sun and in the earth,''
  Phys.\ Atom.\ Nucl.\  {\bf 65 } (2002)  382-391.



\bibitem{jungman}
G.~Jungman, M.~Kamionkowski, K.~Griest,
  %``Supersymmetric dark matter,''
  Phys.\ Rept.\  {\bf 267 } (1996)  195-373.
  [hep-ph/9506380].
\bibitem{Abada}
  A.~Abada, C.~Biggio, F.~Bonnet, M.~B.~Gavela, T.~Hambye,
  %``Low energy effects of neutrino masses,''
  JHEP {\bf 0712}, 061 (2007).
  [arXiv:0707.4058 [hep-ph]].
  \bibitem{Ma}
  E.~Ma, M.~Raidal, U.~Sarkar,
  %``Verifiable model of neutrino masses from large extra dimensions,''
  Phys.\ Rev.\ Lett.\  {\bf 85 } (2000)  3769-3772.
  [hep-ph/0006046].

  \bibitem{Hambye}
  S.~Andreas, T.~Hambye, M.~H.~G.~Tytgat,
  %``WIMP dark matter, Higgs exchange and DAMA,''
  JCAP {\bf 0810 } (2008)  034.
  [arXiv:0808.0255 [hep-ph]].
\bibitem{Hdecay}
  A.~Djouadi, J.~Kalinowski and M.~Spira,
  %``HDECAY: A Program for Higgs boson decays in the standard model and its supersymmetric extension,''
   Comput.\ Phys.\ Commun.\  {\bf 108} (1998) 56  [hep-ph/9704448].  %%CITATION = HEP-PH/9704448;%%

\bibitem{DeltaDECAY}
  M.~Aoki, S.~Kanemura, K.~Yagyu,
  %``Testing the Higgs triplet model with the mass difference at the LHC,''
  [arXiv:1110.4625 [hep-ph]].
  \bibitem{Kopp}
  A.~Esmaili and O.~L.~G.~Peres,
  %``Indirect Dark Matter Detection in the Light of Sterile Neutrinos,''
  arXiv:1202.2869 [hep-ph];
  %%CITATION = ARXIV:1202.2869;%%
 C.~A.~Arguelles and J.~Kopp,
  %``Sterile neutrinos and indirect dark matter searches in IceCube,''
  arXiv:1202.3431 [hep-ph].
  %%CITATION = ARXIV:1202.3431;%%


\bibitem{fit}
C.~Giunti and M.~Laveder,
  %``3+1 and 3+2 Sterile Neutrino Fits,''
  Phys.\ Rev.\ D {\bf 84} (2011) 073008  [arXiv:1107.1452 [hep-ph]];  %%CITATION = ARXIV:1107.1452;%%
C.~Giunti,
  %``Sterile Neutrino Fits,''
   arXiv:1106.4479 [hep-ph];  %%CITATION = ARXIV:1106.4479;%%
B.~Bhattacharya, A.~M.~Thalapillil and C.~E.~M.~Wagner,
  %``Implications of sterile neutrinos for medium/long-baseline neutrino experiments and the determination of $\theta_{13}$,''
  arXiv:1111.4225 [hep-ph];  %%CITATION = ARXIV:1111.4225;%%
   G.~Karagiorgi,
  %``Confronting Recent Neutrino Oscillation Data with Sterile Neutrinos,''
  arXiv:1110.3735 [hep-ph];  %%CITATION = ARXIV:1110.3735;%%
 G.~Karagiorgi,
  %``Confronting Recent Neutrino Oscillation Data with Sterile Neutrinos,''
   arXiv:1110.3735 [hep-ph].  %%CITATION = ARXIV:1110.3735;%%



  \bibitem{vernon}

 V.~Barger, Y.~Gao, D.~Marfatia,
  %``Is there evidence for sterile neutrinos in IceCube data?,''
  [arXiv:1109.5748 [hep-ph]];
  see also, S.~Razzaque and A.~Y.~Smirnov,
  %``Searching for sterile neutrinos in ice,''
  JHEP {\bf 1107} (2011) 084
  [arXiv:1104.1390 [hep-ph]];
  %%CITATION = JHEPA,1107,084;%%
 F.~Halzen,
  %``Sterile Neutrinos and IceCube,''
  [arXiv:1111.0918 [hep-ph]].

\bibitem{Holanda}
P.~C.~de Holanda, A.~Y.~Smirnov,
  %``Solar neutrino spectrum, sterile neutrinos and additional radiation in the Universe,''
  Phys.\ Rev.\  {\bf D83 } (2011)  113011.
  [arXiv:1012.5627 [hep-ph]].
  \bibitem{minos}
P.~Adamson {\it et al.} [ MINOS Collaboration ],
  %``Active to sterile neutrino mixing limits from neutral-current interactions in MINOS,''
  Phys.\ Rev.\ Lett.\  {\bf 107 } (2011)  011802.
  [arXiv:1104.3922 [hep-ex]].
\bibitem{Giunti}
 C.~Giunti, Y.~F.~Li,
  %``Matter Effects in Active-Sterile Solar Neutrino Oscillations,''
  Phys.\ Rev.\  {\bf D80 } (2009)  113007.
  [arXiv:0910.5856 [hep-ph]];
C.~Giunti, Y.~-F.~Li,
  %``Matter Effects in Solar Neutrino Active-Sterile Oscillations,''
  Prog.\ Part.\ Nucl.\ Phys.\  {\bf 64 } (2010)  213-215.
  [arXiv:0911.3934 [hep-ph]].








\end{thebibliography}
\end{document}